Correlation of magnetic moments and angular momenta for stars and planets.

A. Dolginov

Rensselaer Polytechnic Institute, Troy, NY, 12180, USA

Gmail : arkady.dolginov@gmail.com

Abstract

The observed correlation of the angular momenta $L^{ik}$ and magnetic moments $\mu_{lm}$ of celestial bodies (the Sun, planets and stars) was discussed by many authors but without any explanation. In this paper a possible explanation of this phenomenon is suggested. It is shown that the function $\Phi_{lm} = (\eta / r_g) L^{ik} R_{iklm}$ satisfy Maxwell equations and can be considered as a function which determine the electro-magnetic properties of rotating heavy bodies. The $R_{iklm}$ is the Riemann tensor, which determines the gravitational field of the body, $r_g$ is the gravitational radius of the body, and $\eta$ is the constant which has to be determined by observations. The field $\Phi_{lm}$ describe the observed $\mu \rightleftarrows L$ correlation. In particular the function $\Phi_{l0}$ describe the electric field created by rotating heavy bodies. It is possible that the observed electric field of the Earth is created by the Earth rotation.

1. The observed dependence of the magnetic moments on the angular momenta of celestial bodies

The common accepted theory on the origin of the magnetic field of stars and planets is based on the assumption that the field is maintained by hydro-magnetic dynamo. Indeed, it is possible to select the appropriate motions inside the celestial bodies which can support the dynamo action. The assumed motions do not contradict to the existed models of stars and planets. The numerical calculations of the magnetic field confirm the dynamo model. Unfortunately there are no certain evidences on motions inside the bodies and the theory is based on more or less true assumption on these motions. Nevertheless, it seems that the general approach to the problem is true, through some important data remain unexplained. Some problem is connected with the existed correlation of the angular momenta of the rotation and magnetic moments of celestial bodies.

P.M Blackett in 1947, based on the data on Earth, Sun and one other star, speculated that there exists an unknown universal physical law: $(\mu / L) = \beta G^{1/2} / 2c = 4.10^{-15} \beta$ for all astrophysical objects. Here $\mu$ is magnetic moment, $L$ is angular momentum and $\beta$ is of the order of unity. Blackett did not specify what the origin of this law might be. This result was never generally accepted and by the 1950 even Blackett refute this idea.

However, the recent observations show that the $\mu \rightleftarrows L$ correlation exists for planets and for other celestial bodies ( Russell, Dolginov, Arge et al. Cain et al). In this paper we suggest a possible explanation of this phenomenon.

The motions of matter inside different bodies has to be very different. However, observations shows the similar connection of magnetic moments and angular momenta for absolutely different bodies. The observed dependence of the logarithm $\eta = lg(\mu/\mu_0)$ on the $\zeta = lg(L/L_0)$ is presented below (Dolginov 1988):

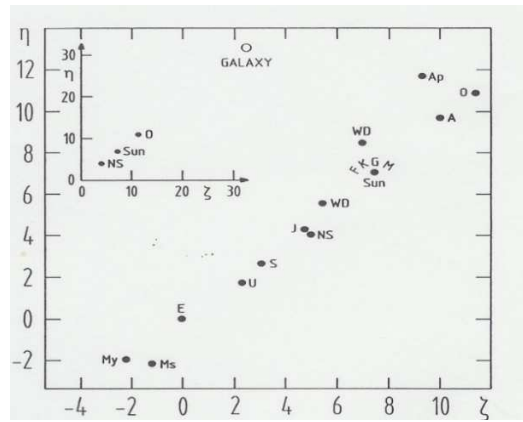

Here $\mu$ is used for the magnetic moments of planets and for the average magnetic momenta of groups of stars, mentioned in this figure, $L$ is the angular momentum. The magnetic moment of the Earth is $\mu_0$, the angular moment is $L_0$. The scattering of these values is very large and it is the reason why the logarithmic scale is used here. There are marked: My-Mercury, E-Earth, Ms-Mars, J-Jupiter, S-Saturn, U-Uranus, N-Neptune, S-Sun. The classes of ordinary stars, used in astronomy, are: (A, O, F, K, G, M). White dwarfs (WD). Neutron Stars (NS). We present here the average values of η and ζ, for stars belonging to each class. The spread of the observed values, within groups of stars of the corresponding class, reaches two orders of magnitude, but almost linear dependence of the mean values extends to twenty orders of magnitude.

The $\mu \rightleftarrows L$ correlation has been investigated in more detail (Arge et al 1995) for 727 astrophysical objects divided into seven groups (solar system, cool stars. hot stars, isolated white dwarfs, cataclysmic variables, isolated pulsars, binary pulsars). In the table presented below the simplified expression for angular momentum $L$ was used: $L = 2\pi M R^2 / P$, where $P = 2\pi/\omega$ is the rotation period. $\mu = B_p r_*^2 / 2$, and $B_p$ is the polar field in Gauss. For the Sun: $B_p \approx 10$ Gauss, $r_*$ is the radius of a body. It is derived: $\log \mu = B \log L + A$. Here $\sigma_B$ is the standard deviation in the slope and in the intercept $\sigma_A$. The $C.C.$ is the correlation coefficient.

| Class | Number | B | $\sigma_B$ | A | $\sigma_A$ | C.C. |
|---|---|---|---|---|---|---|
| Cool stars | 54 | 1.091 | 0.062 | 0.76 | 0.566 | 0.925 |
| Solar system | 8 | 0.924 | 0.083 | -0.645 | 0.349 | 0.976 |
| Hot stars | 171 | 0.571 | 0.058 | 5.442 | 0.616 | 0.602 |
| (Hot stars-SG) | -167 | -0.383 | -0.05 | -7.355 | -0.528 | -0.51 |
| Isolated white dwarfs | 14 | -0.119 | 0.355 | 8.731 | 2.004 | -0.096 |
| Isolated pulsars | 429 | -0.559 | 0.061 | 7.115 | 0.328 | -0.405 |
| Cataclysmic variable | 19 | -1.008 | 0.097 | 14.481 | 0.676 | -0.93 |
| Binary pulsars | 32 | -1.167 | 0.04 | 9.613 | 0.221 | -0.983 |

The plotted points span a range of some 20 orders of magnitude in both $\mu$ and $L$, and much more if we include the Galaxy as a whole. It was shown that the center of distribution for all these sub-samples, excluding binary pulsars and cataclysmic variable, are located close to the same line: $\lg \mu = (1.294 \pm 0.017) \lg L - 1.845$ with the correlation coefficient: $0.941$.

We will consider below only the case of single stars. The case of binary stars needs a separate investigation. Our result also describes approximately the $\mu \rightleftarrows L$ correlation for most planets in solar system. However, it needs a special investigation for such bodies as the Moon, which rotation is strongly connected with their orbital motion. More detail investigation of $\mu \rightleftarrows L$ correlation for binary and multibody systems is necessary to decide how their magnetic field is determined by gravitation, by the spin and by the orbital angular momentum of the body

2. Possible explanation of $\mu \rightleftarrows L$ correlation

To explain the $\mu \rightleftarrows L$ correlation we assume: (a) existence of electromagnetic field which is created by rotation of a body (b) this field has to be described by Maxwell equation (c) this field has to be linearly dependent on the body angular momentum (d) this field is significant only for very heavy bodies such as stars and planets. To this purpose consider the anti-symmetric second rank tensor:

$$\Phi_{lm} = (\eta / r_g) L^{ik} R_{iklm} \qquad (1)$$

Here $R_{iklm}$ is the Riemann tensor which determine the gravitational field of the body, $r_g = 2Gm/c^2$ is the gravitational radius of the body and $\eta$ is the constant which has to be determined by observations.

$$R_{iklm} = \frac{1}{2}\{\frac{\partial^2 g_{im}}{\partial x^k \partial x^l} + \frac{\partial^2 g_{kl}}{\partial x^i \partial x^m} - \frac{\partial^2 g_{il}}{\partial x^k \partial x^m} - \frac{\partial^2 g_{km}}{\partial x^i \partial x^l}\} + g_{np}\{\Gamma^n_{kl}\Gamma^p_{im} - \Gamma^n_{km}\Gamma^p_{il}\}$$

$$= \frac{4\pi G}{c^4}\{g_{km}T_{il} - g_{im}T_{kl} + g_{il}T_{km} - g_{kl}T_{im} - \frac{2}{3}(g_{il}g_{km} - g_{kl}g_{im})T\} + (R_{iklm})_0$$

$$= \frac{1}{2}\{R_{il}g_{km} - R_{im}g_{kl} + R_{kl}g_{im} - R_{km}g_{il} - \frac{1}{3}(g_{il}g_{km} - g_{im}g_{kl})R\} + C_{iklm} \quad (2)$$

Here $(R_{iklm})_0$ is the vacuum part of the Riemann tensor and $C_{iklm}$ is the Weyl tensor.

Angular momentum $L^{ik}$ depends on the body characteristics, but not on the space-time coordinates $x^i$ and $t$:

$$L^{ik} = \int (x^l dP^K - x^k dP^i) \quad (3)$$

Using Bianchi identities and contracted Bianchi identities:

$$R_{nmik;l} + R_{nkil;m} + R_{nkmi;l} = 0 \quad \text{and} \quad R^{nmik}_{;l} = R^n_{k;m} - R^m_{k;n} \quad (4)$$

where $R^n_k$ is the Ricci tensor and $R^n_{k;m}$ is the covariant derivative of $R^n_k$ we obtain Maxwell equations for $\Phi^{nm}$:

$$\Phi_{ik;l} + \Phi_{li;k} + \Phi_{kl;i} = 0 \qquad \Phi^{nm}_{;n} = J^m \quad (5)$$

$$J^m = (\eta/r_g)L^{ik}(R^m_{i;k} - R^m_{k;i}) = (8\pi G/c^2)(\eta/r_g)L^{ik}(T^m_{i;k} - T^m_{k;i}) \quad (6)$$

where $J^m$ determine the effective current, created by the body rotation. This current is divergence free

$$J^m_{;m} = (\eta/r_g)L^{ik}(R^m_{i;k} - R^m_{k;i})_{;m} = 0 \quad (7)$$

The result (7) is not identical to the four dimensional divergence low for the energy-momentum tensor: $T^m_{i;m} = 0$. It was proved (Pagels, 1961) that $(R^m_{i;k} - R^m_{k;i})_{;m} = 0$. This is an additional condition that follows from the Einstein equation.

The so called "electromagnetic components of the Riemann tensor" were considered in many papers (see, for example, the paper of B. Mashhoom and references there) but without any connection to problems of $\mu \rightleftarrows L$ correlation for astrophysical objects. The main assumption of our paper is the assumption that the field, which is described by $\Phi_{lm}$ is a real electromagnetic field which interacts with electrical charges. This assumption does not contradict observations because the effect of this field can be observed only for very heavy objects, such as stars and planets. The gravitation was never considered in connection with the problem of $\mu \rightleftarrows L$ correlation.

Let us consider the simplest case of a heavy spherical body having the mass $m$, radius $r_s$ and constant mass density $\rho$. The metric of the body gravitational field is determined from:

$$ds^2 = (1+2\varphi/c^2)c^2\,dt^2 + (-1+2\varphi/c^2)(dx^2 + dy^2 + dz^2) \qquad (8)$$

We will consider the case of a weak field and use the linear approximation. Inside the body: $\varphi = -2\pi\rho G(r_s^2 - r^2/3)$. In this case the nonzero $R_{iklm}$ are:

$$R_{1212} = R_{1313} = R_{2323} = -r_g/r_s^3, \qquad R_{iklm} = R_{lmik} = -R_{kilm} \qquad (9)$$

The metric of rotating bodies depends on angular momentum (the Kerr metric, for example). However, the dependence of the metric $g_{ik}$ on the potential $\varphi$ is much stronger than on $L^{ik}$. Using the obtained $R_{iklm}$ values and choosing the axis Z along the angular momentum, i.e. taking $L^{IK} = L^{12} = L_Z = I\Omega$ where $I$ is the moment of inertia and $\Omega$ is the angular velocity, we obtain the field inside the body that is approximately constant. Gravitational potential outside the body is: $\varphi = -GM/r$. In this case:

$$R_{1212} = (r_g/2r^3)(1-3Cos^2\theta), \quad R_{1223} = (3r_g/2r^3)Cos\theta Sin\theta Sin\varphi, \qquad (10)$$

$$R_{1213} = (3r_g/2r^3)Cos\theta Sin\theta Cos\varphi$$

Using (1) and (8) we can see that outside the body the $\Phi_{lm}$ represents a dipole field:

$$\Phi_{23} = (3\eta/2r^3)I\Omega Cos\vartheta Sin\vartheta Cos\varphi \to H_x,$$

$$\Phi_{12} = (\eta/2r^3)I\Omega(1-3Cos^2\theta) \to H_z,$$

$$\Phi_{31} = (3\eta/2r^3)I\Omega Cos\vartheta Sin\vartheta Sin\varphi \to H_y, \qquad (11)$$

We can see from (5)-(11) that there exists an antisymmetric tensor $\Phi_{lm}$ that satisfy Maxwell equations and is determined by the divergent free current. Can the function $\Phi_{lm}$ determine the interaction of charged particles? In other words: is $\Phi_{lm}$ the tensor of electromagnetic field, created by rotation of heavy bodies? This possibility does not follow from the $\Phi_{lm}$ definition, but there are no arguments to exclude it. This possibility can be proved or denied only by experiment or astrophysical observations. The estimate value of the $\eta$ is of the order of $10^{-15} cm^{1/2} g^{-1/2}$ that is approximately equivalent to $G^{1/2}/2c$.

Astrophysical objects have both-the poloidal and toroidal magnetic fields. The toroidal field is easily created from poloidal field by differential rotation of the convective matter inside the body. Such rotation inside the body is common for most astrophysical objects, but the creation of the poloidal field needs complicated cyclonic motions.

The electric field is expressed by $\Phi_{l0} \to E_l$. If the interval $ds$ is determined by (8) then the electric field of a body is not zero only if the gravitational potential $\varphi$ depends on time. For example, time dependent are tidal forces. The stationary electric field is created in the system if the metric of the system contains terms with $g_{mn}$ where $m \neq n$.

It is known that the significant electric field exists in the Earth atmosphere. The total electric current reaching the earth surface at any time is approximately constant. The Earth surface has the positive charge of 1800 amperes, 400000 volts – a power of 700 megawatts (Feynman 1957). What maintains the voltage? There are many attempts to describe this field as due to ionization of atmosphere by cosmic rays, by solar wind etc., but there is no commonly accepted explanation. Regarding other celestial bodies, there is no observations related to their global electric field. The existed electrical field on the Earth was also not observed from cosmos.

The magnetic energy of the body is quadratic with respect to $\Phi_{lm}$, and, therefore, the magnetic energy does not depend on the direction of angular momentum. In other words, we have *degenerated state*. In such a case, some perturbations may change the magnetic field direction with respect to the direction of the angular momentum $L^{ik}$. The change of the field polarity may be due to some irregular or regular changes of the internal motions inside the body. The change of the field direction took place several times during the Earth history.

The field of white dwarfs and neutron stars is usually explained as a relict field conversed during the star contraction. If the initial surface field of a star with the radius $r_s$ was $H_{in}$, then the field after contraction to the radius $r_f$ will be $H_f = H_{in}(r_f/r_{in})^2$. It was not taking into account that some fraction of mass is lost in the process of the star contraction and the conductivity could be significantly changed. The conductivity of compact objects is provided by electron transfer, whereas in non-compact parent star it was provided by turbulent plasma motions. It seems that there is no a priory reason to think that the centers of distribution of compact objects in an $\mu \rightleftarrows L$ plot should have any relation to the centers of distribution of non-compact objects. However, the analysis (see Arge et al.1995, and Dolginov 1989) demonstrates that the centers of the $\mu \rightleftarrows L$ distribution for white dwarfs and neutron stars are located on the same curve as for all other stars and planets. It can be explained if there exist some kind of universal $\mu \rightleftarrows L$ connection also for these stars. The possible explanation of these observations is the existence of the field $\Phi_{lm}$. The difference of white drafts and neutron stars from other stars may be connected with the white dwarfs and neutron stars internal structure. The table of $\mu \rightleftarrows L$ presented above do not include any data related to the inner structure and gravitation of astrophysical objects. In reality these structures are very complicated, including differential rotation of different layers, convective motions etc. These motions are considered in the dynamo theory, which is the most accepted theory of the magnetic field generation in stars and planets. Motions inside the bodies could be important not only for dynamo actions but also for the $\mu \rightleftarrows L$ correlation.

Unfortunately, there are no direct evidences on the internal structures of celestial bodies, though some qualitative estimations are apparently true. The Earth solid inner core, the liquid core and the mantle are rotating with slightly different angular velocities, and have the axis of rotation slightly inclined to each other. Interiors of white dwarfs and neutron stars is much more inhomogeneous than the Earth interior. In particular, the shift of the point for isolated white dwarfs in the table presented above, could be due to strong gravitational field of white dwarfs, which is described by the term $R_{iklm}$ in the formula (1) and to the different rotation of the different internal layers which was not taken into account in the presented table.

The internal structures of massive white dwarfs and neutron stars contain superfluid layers which provide different rotation of different internal regions. These layers can rotate independent on the matter on the star surface. It is possible that the magnetic field of neutron stars is determined mostly by the rotation of the star interior, which is much heavier than the surface layers and could rotate opposite to the surface layers. This may explain the negative sign of the coefficient $B$ in the table presented above.

The existence of the motions inside celestial bodies which create the dynamo action does not contradict to the possibility of the universal field created by the body rotation.

It was shown by L.Ferarro, et al (2015) that : (1)."Highly magnetic white dwarfs tends to exhibit a complex and non-dipolar field structure with some objects showing the presence of higher order of multipoles. (2). There is no evidence that fields of highly magnetic white dwarfs decay over time, which is considered with the estimated ohmic decay time scales of $10^{11}$ years."

This result can be explained taking into account that the ohmic decay do not destroy the field $\Phi_{lm}$ created by the body rotation.

3. Conclusion

Astronomical data show that there exists correlation of the magnetic moments and angular momenta for all celestial bodies. We assume that rotation of heavy bodies create the electromagnetic field $\Phi_{lm} = (\eta / r_g) L^{ik} R_{iklm}$. It can explain the observed correlation. The component $\Phi_{10}$ describes the electric field of rotating body. It could explain the observed electric field on the Earth. The obtained results do not contradict to possibility of the dynamo action inside astrophysical objects.

References

1.Arge C,N.,et al. (1995), Astrophys.Journ., 443,795

2.Blackett P.M.S. (1952), Philos.Trans..Roy.Soc., London, A 245,309

3.Cain J.C. et al. (1995),Journ.Geophys.Res., 100,9439

4.Dolginov A.Z. (1988),Phys.Rep. 163, N 6, 337

5.Dolginov A.Z. (1989), in galactic and Intergalactic Magnetic Fields ed.R

Beck et al.(Dordrecht: Kluwer) 27

6.Feynman R.   (1957), Lectures on Phys. V2, ch.9, Pergamon Press, London

7.Ferarro L. et al (2015), arXiv:astro-ph.:1504,08072, v1

8. Mashhoon B. (2000), arXiv: gr-qs.:0011014 v1

9.Pagels H. (1963), Journ. Math. Phys. 4, 731

10. Russell C.T. (1979), Nature, 281,552